\documentclass{article}

\usepackage{amsmath,amsfonts,amsthm}
\usepackage{fullpage}
\usepackage{qcircuit}
\usepackage{xspace}
\usepackage[usenames,dvipsnames]{xcolor}
\usepackage{listings}
\usepackage{subcaption}
\usepackage{hyperref}
\usepackage{cleveref}
\usepackage{tikz}
\usepackage{mathtools}
\usepackage{booktabs}
\usepackage[normalem]{ulem}

\usepackage{braket}
\usepackage{authblk}

\theoremstyle{plain}
\newtheorem{theorem}{Theorem}[section]

\theoremstyle{definition}

\theoremstyle{remark}
\newtheorem{remark}[theorem]{Remark}
\newtheorem*{notation*}{Notation}

\newcommand{\name}{\texttt{staq}\xspace}
\newcommand{\Name}{\texttt{staq}\xspace}

\newcommand{\edit}[2]{#2}

\usetikzlibrary{shapes,arrows,positioning,matrix,fit}
\tikzset{
>=stealth',
  def/.style={
    rectangle, 
    rounded corners,
    draw=black,
    minimum height=3em, 
    text centered},
  node distance=1cm,
}

\title{\Name -- A full-stack quantum processing toolkit}
\author[1,2]{Matthew Amy\thanks{matt.amy@dal.ca}}
\author[1,3,4]{Vlad Gheorghiu\thanks{vlad@softwareq.ca}}
\affil[1]{softwareQ Inc., Kitchener ON, Canada}
\affil[2]{Department of Mathematics \& Statistics, Dalhousie University, Halifax NS, Canada}
\affil[3]{Institute for Quantum Computing, University of Waterloo, Waterloo ON, Canada}
\affil[4]{Department of Combinatorics and Optimization, University of Waterloo, Waterloo ON, Canada}

\date{Version of \today}

\begin{document}
\maketitle

\begin{abstract}
    We describe \name, a full-stack quantum processing toolkit written in standard C++. \Name is a quantum compiler toolkit, comprising of tools that range from quantum optimizers and translators to physical mappers for quantum devices with restricted connectives. The design of \name is inspired from the UNIX philosophy of ``less is more", i.e. \name achieves complex functionality via combining (piping) small tools, each of which performs a single task using the most advanced current state-of-the-art methods. We also provide a set of illustrative benchmarks.
\end{abstract}

\tableofcontents

\section{Introduction\label{sct:intro}}
Quantum computing is a new paradigm of physics that promises significant computational advantages for a plethora of applications, ranging from optimizing, material design, drug discovery to sensing and measurement and secure communication. The idea of harnessing the power of quantum mechanics to perform computations that are believed to be much harder or even intractable by classical computers dates back to Feynman~\cite{feynman-82}. However, due to experimental challenges, the first public-access small programmable quantum computing platforms appeared within the last five years, whereas the first demonstration of a computational task that can be performed significantly faster by a quantum computer was publicly released on October 2019~\cite{Arute:2019aa}.

The currently available quantum computing platforms are  not (yet) fault tolerant, i.e. they can not perform arbitrary long quantum computations with arbitrarily low error, but consist mainly of noisy qubits with restricted connectivity, for which the computation length is restricted by the depth of the logical circuit to be run.
Such platforms are informally termed ``Noisy Intermediate-Scale Quantum" computers (NISQ)~\cite{p18}, and represent the first step towards realization of large-scale fault-tolerant quantum platforms.

Quantum algorithms are usually described in a high-level language (e.g. plain English or quantum ``pseudo-code") then ``translated" into a quantum circuit consisting of a series of quantum gates applied in a sequential manner\footnote{The quantum gate model is not the only quantum computing model, and other\edit{}{s} exist, such as Measurement-Based Quantum Computing~\cite{PhysRevLett.86.5188}, Adiabatic Quantum Computing~\cite{RevModPhys.90.015002}, etc. However, in this paper, we will focus on the gate model, as currently it seems to be the favourite among the community that believes in full-scale fault-tolerant quantum computation.}, followed by a measurement from which the end result of the algorithm is being inferred via classical post-processing techniques, see e.g.~\cite{NC00} for more details.

The translation from a high-level quantum algorithm to a quantum circuit is informally called ``quantum compiling", and consists of a series of steps, which depend on the particular quantum architecture that is being used. Those steps are usually thought of as a quantum compiling top-down ``stack" (with the most abstract layers higher up in the stack), and may involve e.g. translation of parts of a high level quantum algorithm to logical Boolean circuits (such as the mapping of quantum oracles in quantum searching algorithms to Boolean functions), converting Boolean functions to quantum reversible circuits, optimizing the latter in terms of a particular cost and taking into account connectivity constraints (if any), and finally mapping the resulting quantum circuit to a specific physical architecture, or translating it to some particular kind of quantum machine assembly language.

Any optimizations or improvements along the stack affect (beneficially) the quantum computation speed (QPU cycles/wall time), and may even allow longer computations on NISQ devices, which otherwise would be \edit{in-feasible}{infeasible} due to their prohibitively-large circuit depth. Therefore constructing ``full-stack" quantum processing toolkits is of paramount importance for both the NISQ regime and also far-future large-scale fault-tolerant quantum platforms.

\name represents a joint effort at \href{https://www.softwareq.ca}{softwareQ Inc.} to construct such a full-stack quantum computing toolkit. Our effort is not the first and likely not the last in the fast dynamic field of quantum software\edit{}{~-- the myriad of existing quantum compilers and toolkits includes Quipper \cite{glrsv13}, ScaffCC \cite{jpkhlcm15}, Quil/PyQuil \cite{scz16}, ProjectQ \cite{sht18}, $Q\ket{SI}$ \cite{lwglhdy17}, Q\# \cite{sgtaghkmpr18} and Strawberry Fields \cite{kiqbaw18} to name just a few}. \edit{What differentiates \name from the other quantum software stacks is the minimalist design, inspired from the UNIX~\cite{Ritchie:1974:UTS:361011.361061} world, in which one can achieve complex functionality via combining (piping) small tools, each performing one single task well. \name is a collection of such tools, ranging from circuit optimizers and translators to physical mappers for NISQ architectures with restricted connectives. Our design consideration is in high contrast with monolithic software design, in which functionality is being attained via a single tool that does ``everything". Our modular design reduces software maintenance efforts, increases backwards compatibility for future releases, and allows adding new functionality relatively easy.}{In contrast to those compilers, \name~-- inspired by the UNIX~\cite{Ritchie:1974:UTS:361011.361061} philosophy of building up complex functionality by combining (piping) small command-line tools -- is designed as a collection of minimalist tools, ranging from circuit optimizers and translators to physical mappers for NISQ architectures with restricted connectives, operating on a single textual language, openQASM~\cite{cbsg17}. While modularity in compilers is hardly a novel concept, \name is differentiated by a Clang~\cite{clang} style approach of representing programs by, and operating directly on, syntax trees, rather than various internal representations. This approach allows us to export compiler modules directly as single-purpose source-to-source tools that make minimal, independent changes to a given program, and which can then be combined at will with other tools or compilers.}

\edit{In addition}{As a compiler and software toolkit}, \name~\edit{uses }{integrates \emph{in a production setting }}the latest state-of-the-art methods in quantum compiling \edit{while}{} targeting the whole spectrum of the quantum software stack, starting from the abstract higher algorithmic layers and ending at the physical mapping layer. \edit{}{In particular, \name includes state-of-the-art techniques which have either never been implemented, or otherwise only implemented in restricted academic settings. }Moreover, \name is highly portable, being written in standard C++ (using the C++17 standard), and fast, as shown by our benchmarks in~\Cref{sec:perf}. \Name also offers some unique features, such as the ability \edit{of translating logical Boolean circuits specified in an industrial language such as Verilog~\cite{1620780} to reversible circuits}{to use and synthesize classical logic, specified in an industrial-strength language such as Verilog~\cite{1620780}, directly within openQASM code}. Finally, \name can generate quantum code in a variety of formats that encompass \edit{most}{many} of the currently available quantum platforms~\cite{cbsg17,cirq,Qiskit,scz16,sht18,sgtaghkmpr18}.

The rem\edit{}{a}inder of this paper describes the \name toolkit and its functionality and uses cases while providing a set of benchmarks (\Cref{sct:main}), followed by conclusions and future directions (\Cref{sct:concl}).

\section{\Name\label{sct:main}}

\Name is a new compiler and software toolkit for the openQASM language~\cite{cbsg17} written in C++. The primary goal is to provide a suite of transformation, optimization and compilation tools that can operate on a single, common language, and output to a number of different simulation and hardware execution platforms. On the technical side, a focus of \name is to support state-of-the-art circuit transformation algorithms, which are typically implemented on small subsets of circuits or in restricted research contexts, and apply them natively to \emph{any valid quantum program}. This algorithmic focus is distinct from other quantum computing toolchains, which are typically slow to adopt bleeding-edge techniques. \edit{In this section we give an overview of the architecture, use, and algorithmic methods of \name.}{We first give a brief review of the openQASM language, with the remainder of this section giving an overview of the architecture, usage, and algorithmic methods of \name.}

\paragraph{The openQASM language} The official specification of openQASM can be found in \cite{cbsg17}, but we provide a brief overview here. Programs in openQASM are structured as sequences of declarations and commands. As an intermediate- to low-level language, openQASM provides a small number of basic programming features: declaration of static-size classical or quantum registers, definition of (unitary) circuits or \emph{gates}, gate application, measurement and initialization of qubits, and finally classically controlled gates. The listing below gives an example of an openQASM program performing quantum teleportation:
\begin{lstlisting}
OPENQASM 2.0;
include "qelib1.inc";

gate bellPrep x,y {
  h x;
  cx x,y;
}

qreg q[3];
creg c0[1];
creg c1[1];

bellPrep q[1],q[2];
cx q[0],q[1];
h q[0];
measure q[0] -> c0[0];
measure q[1] -> c1[0];
if(c0==1) z q[2];
if(c1==1) x q[2];
\end{lstlisting}


\subsection{Overview}\label{sec:overview}

To support the minimalist philosophy of small, single-function tools, \name was designed from the bottom-up to allow the manipulation, transformation, compilation and translation of QASM files according to the following goal:
\begin{quote}
	\emph{no process should affect the original structure of the program more than absolutely necessary}.
\end{quote}
In particular, an un-transformed program should output to something that looks identical to the input source code, modulo changes in whitespace. Similarly, one should be able to optimize programs without disturbing the structure of the original program, or to the extent that the developer wishes to enable further optimizations, for instance by first inlining and then performing \emph{whole-program} optimization. 

To achieve this, \name stores and operates \emph{directly on QASM syntax trees}, rather than an intermediate representation. This approach was inspired by Clang~\cite{clang}, which acts as an effective middle-end for the analysis and transformation of C code. In keeping with the Clang style of program analysis, \name provides a powerful set of \emph{Visitor} classes for performing different types of traversal as well as AST splicing, on which all of our transformations are built.

\begin{figure}
\centering
\begin{tikzpicture}
     \node[def] (one) {\begin{tabular}{c} QASM \\ 
	{\small w/ uniform gates, oracles,} \\ {\small ancillas, gate declarations} \end{tabular}};
     \path (one) edge[loop right] node {semantic checker} (one);
     \node[def, below left=1cm and 2.3cm of one] (two)      {\begin{tabular}{c} QASM \\ 
	{\small w/ oracles, ancillas,} \\ {\small gate declarations} \end{tabular}};
     \draw[->, thick] (one) -- node[above left] {desugaring} (two);
     \node[def, right=3cm of two] (three)      {\begin{tabular}{c} QASM \\ 
	{\small w/ ancillas,} \\ {\small gate declarations} \end{tabular}};
     \draw[->, thick] (two) -- node[above] {logic synthesis} (three);
     \node[def, right=4cm of three] (four)      {\begin{tabular}{c} QASM \\ 
	{\small w/ ancillas,} \\ {\small gate declarations} \end{tabular}};
     \draw[->, thick] (three) -- node[above] {\begin{tabular}{c} optimization \\ 
	(simplification, rotation \\ folding, \edit{gray synthesis}{CNOT resynthesis})\end{tabular}} (four);
     \node[def, below =1cm of four] (five)      {\begin{tabular}{c} QASM \\ 
	{\small inline} \end{tabular}};
     \draw[->, thick] (four) -- node[left] {inlining} (five);
     \node[def, left=5.3cm of five] (six)      {\begin{tabular}{c} QASM \\ 
	{\small mapped} \end{tabular}};
     \draw[->, thick] (five) -- node[above] {hardware mapping} (six);
     \node[def, below left=2cm and 4.5cm of six] (o1)      {QASM};
     \node[def, right=1cm of o1] (o2)      {Quil};
     \node[def, right=1cm of o2] (o3)      {ProjectQ};
     \node[def, right=1cm of o3] (o4)      {Q\#};
     \node[def, right=1cm of o4] (o5)      {Cirq};
     \node[def, right=1cm of o5] (o6)      {Resource counts};
     \draw[->, thick] (six) [out=250, in=90] to  (o1);
     \draw[->, thick] (six) [out=260, in=90] to  (o2);
     \draw[->, thick] (six) [out=270, in=90] to  (o3);
     \draw[->, thick] (six) [out=270, in=90] to  (o4);
     \draw[->, thick] (six) [out=280, in=90] to  (o5);
     \draw[->, thick] (six) [out=290, in=90] to  (o6);
     \node[draw,dotted,thick,fit=(o1) (o2) (o3) (o4) (o5) (o6),label={[above right=0cm and 6cm of o6]compilers}] {};
  \end{tikzpicture}
\caption{Overview of the \name toolchain}\label{fig:overview}
\end{figure}
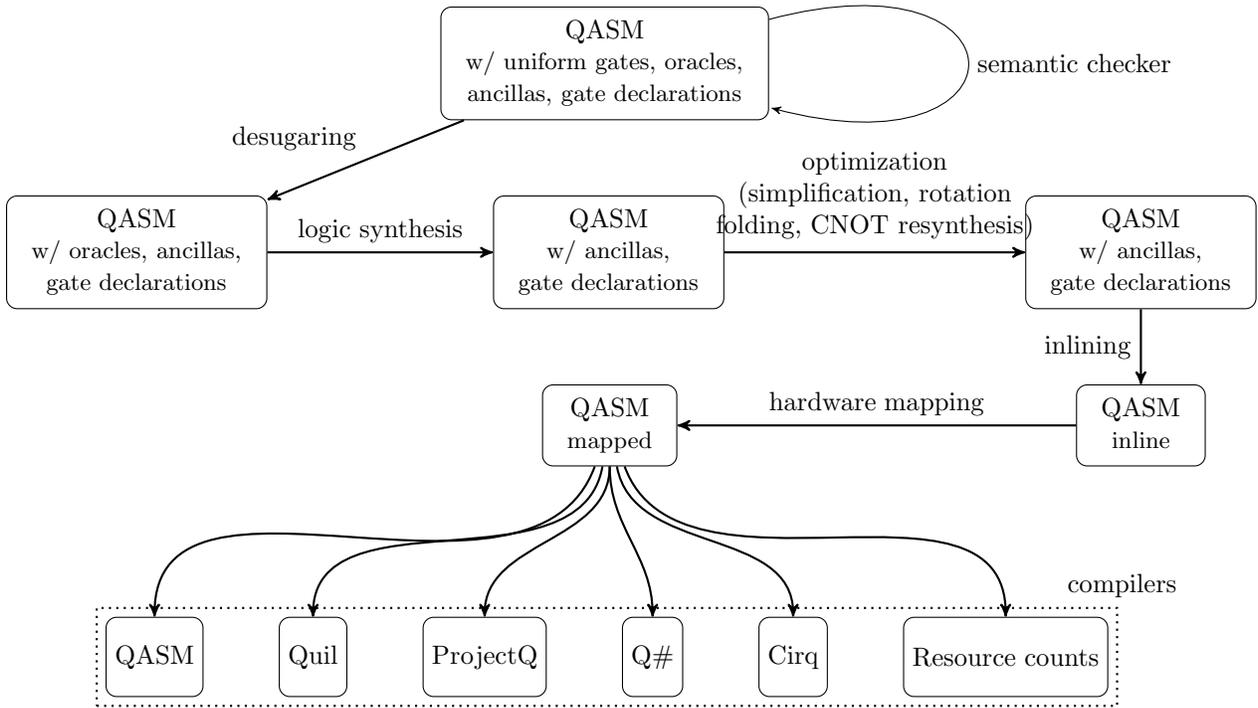

\Cref{fig:overview} gives an overview of the \name toolchain and typical usage. The main command line compiler -- \texttt{\name} -- offers a flexible \emph{pass registration} system, whereby passes are given via command-line arguments and are executed in the order given. In particular, it is often useful to perform basic gate simplifications both before and after other optimizations, or to inline certain gates (e.g., \texttt{ccx}) first, optimize, then inline fully to primitive gates -- these types of usage patterns are supported by the pass registration system, for instance with\footnote{This sequence of passes is also given the explicit name \texttt{-O1}, in analogy to the basic optimization level of GCC.} \texttt{{\name} -s -r -s circuit.qasm} in the former case.

As all of the transformations are defined directly on QASM ASTs, the order of operations is generally interchangeable, with two major exceptions:
\begin{enumerate}
	\item desugaring must occur before any other transformation, and
	\item the program must be \emph{fully} inlined before mapping.
\end{enumerate}
For this reason, the compiler automatically applies a desugaring pass after parsing and semantic checks, and an inlining pass preceding a hardware mapping pass. Desugaring mainly involves replace \emph{uniform gates} -- gates applied to registers -- with a sequence of gates applied to individual qubits. For instance, if \texttt{x} and \texttt{y} are qubit registers of length $2$, the desugarer will replace \texttt{cx x,y;} with \\
\centerline{\texttt{cx x[0],y[0];}}\\
\centerline{\texttt{cx x[1],y[1];}} \\
The semantic checker ensures that all such uniform gates are well-formed according to the specification in \cite{cbsg17}, as well as other semantic properties such as correct argument types.

The general inline pass supports \emph{overrides}, whereby the user can specify which gates should \emph{not} be inlined. By default, the gates defined in \texttt{qelib1.inc} \cite{cbsg17} are not inlined, except before hardware mapping. The remaining synthesis, optimization, and mapping passes are described in more detail in the follow sections.

\paragraph{Tool suite}

In addition to a single compiler, the \name software package also includes a suite of light-weight command line tools which can be chained together using Unix-style \emph{pipelines} to perform a range of compilation tasks. Each tool reads a QASM file from \texttt{stdin}, performs a specific function, and outputs the transformed QASM source on \texttt{stdout} -- as an exception, the \emph{compiler} tools output the QASM source in various other languages. This offers a more flexible and customizable compilation pipeline at the expense of extra parsing stages, as well as the option to only build the relevant tools for a particular use case. For a full description of the available tools, the reader is directed to the \name wiki\footnote{\href{https://github.com/softwareQinc/staq/wiki/The-staq-tool-suite}{https://github.com/softwareQinc/staq/wiki/The-staq-tool-suite}.}.

\paragraph{openQASM extensions}

\Name supports a number of extensions to the openQASM language, both implemented and planned. In particular, \name supports the declaration and use of ancillas local to gate declarations, as well as the declaration of quantum oracles from classical Verilog logic files. These extensions are described in more detail in \Cref{sec:synthesis}. Future planned versions will support iteration and register arguments to gates \`{a} la metaQASM \cite{a19}.

\subsection{Circuit synthesis}\label{sec:synthesis}

A unique feature of \name is the ability to splice classical logic \emph{directly into quantum programs}, and moreover the ability to \emph{synthesize} a circuit implementing the classical logic during compilation. This is done through a QASM language extension adding \emph{oracle} gate declarations, with synthesis handled by the EPFL Logic Synthesis Libraries \cite{EPFLLibraries}. At this time, \name supports classical logic written in (the combinational subset of) the Verilog hardware description language.

To declare an oracle gate in a \name-QASM file, the keyword \texttt{oracle} is used in place of \texttt{gate}, and the classical logic defining the gate is given in the body as the name of a verilog file:
\begin{lstlisting}
oracle MUX sel,x,y,out { "mux.v" }
\end{lstlisting}
The listing below shows the corresponding Verilog file \texttt{mux.v}.
\begin{lstlisting}
// mux.v
module top (a, b, c, d);
  input a,b,c;
  output d;
  wire tmp1,tmp2,tmp3;
  assign tmp1 = a & b;
  assign tmp2 = ~a & c;
  assign tmp3 = ~tmp1 & ~tmp2;
  assign d = ~tmp3;
endmodule
\end{lstlisting}
Combinational logic can be written in Verilog as a sequence of assignments of logical expressions to either \emph{outputs} or temporary \emph{wires}. For a full overview of the Verilog programming language, the reader is directed to \cite{tm13}. Due to the reversibility of quantum oracles, there must be exactly one oracle input for every input and output of the given Verilog file. Oracle inputs are mapped to Verilog inputs and outputs sequentially, regardless of naming.

The logic synthesis pass of the compiler visits the AST and replaces each oracle declaration with a corresponding gate declaration. When an oracle declaration is encountered, it parses the specified file to generate an Majority-inverter graph (MIG), which is then synthesized over the Clifford group and $T$ gates using the EPFL implementation of the LUT-based Hierarchical Reversible Logic Synthesis (LHRS) framework \cite{srwm19}. 

In general, a classical function may require \emph{ancillas} to be implemented reversibly, and so synthesis of classical logic may require additional ancillas that have not been given directly as inputs to the gate. \Name handles the introduction of ancillas with another QASM language extension adding support the declaration of local ancillas within gate bodies. Inside gate declarations, both \emph{clean} and \emph{dirty} ancilla registers -- registers initialized in the state $\ket{00\cdots 0}$ or in some unknown state, respectively -- can be declared similarly to regular QASM registers by using the keyphrases \texttt{ancilla} or \texttt{dirty ancilla}, respectively. All ancillas are assumed to be returned to their initial state at the end of the gate body, and it remains the programmer's responsibility to ensure that this requirement is satisfied. Lightweight verification methods such as \emph{path-sum verification} \cite{a18} can be adopted in the future to ensure that all ancillas are properly cleaned at the end of a gate.

The result of synthesizing the \texttt{MUX} gate above is shown below. Despite the use of temporary wires in the input Verilog file, the LHRS synthesis algorithm is able to find an ancilla-free implementation, hence the resulting local ancilla register has size $0$.

\begin{minipage}{0.49\textwidth}
\begin{lstlisting}
gate MUX sel,x,y,out {
	h out;
	cx x,out;
	tdg out;
	cx sel,out;
	t out;
	cx x,out;
	tdg out;
	cx sel,out;
	t out;
	cx sel,x;
	tdg x;
	cx sel,x;
	t sel;
	t x;
	h out;
---------\\----------
\end{lstlisting}
\end{minipage}
\begin{minipage}{0.49\textwidth}
\begin{lstlisting}
---------\\----------
	h out;
	cx y,out;
	t out;
	cx sel,out;
	t out;
	cx y,out;
	tdg out;
	cx sel,out;
	tdg out;
	cx sel,y;
	tdg y;
	cx sel,y;
	t sel;
	tdg y;
	h out;
}
\end{lstlisting}
\end{minipage}

\begin{remark}
The synthesized circuit above has $T$-count $14$, since the 3-qubit multiplexor can be implemented with $2$ Toffoli gates. This can further be reduced to $8$ with the light optimization pass, \texttt{-O1}.
\end{remark}


\paragraph{Ancilla management}
As local ancilla allocation is non-standard QASM and moreover not supported by most QPUs, \name performs automatic ancilla management during the inlining phase of compilation. In particular, local ancilla declarations are hoisted, as regular qubit registers, to the top of the global scope when a gate is inlined.

Since ancillas are assumed to be returned \emph{clean} -- i.e. returned in their initial state -- it is not always necessary to allocate a new ancilla for every gate application. \Name handles the re-use of ancillas during compilation by maintaining a pool of previously allocated ancillas and available dirty qubits, re-using qubits from these pools to fulfill ancilla requirements whenever possible. \Cref{fig:ancillamanagement} shows an example of \edit{ancilla}{ both dirty and clean ancilla allocation and} sharing between gate applications.

\begin{figure}
\centering
\begin{subfigure}[b]{0.49\textwidth}
\centering
\begin{lstlisting}
OPENQASM 2.0;
include "qelib1.inc";

gate foo a {
  ancilla b[1];
  cx a,b[0];
}

gate bar a {
  ancilla b[1];
  dirty ancilla c[1];
  cx a,b[0];
  cx a,c[0];
}

qreg x[2];
foo x[0];
bar x[0];


\end{lstlisting}
\caption{Before inlining}
\end{subfigure}
\begin{subfigure}[b]{0.49\textwidth}
\centering
\begin{lstlisting}
OPENQASM 2.0;
include "qelib1.inc";

gate foo a {
  ancilla b[1];
  cx a,b[0];
}

gate bar a {
  ancilla b[1];
  dirty ancilla c[1];
  cx a,c[0];
  cx c[0],b[0];
}

qreg anc[1]
qreg x[2];
cx x[0],anc[0];
cx x[0],x[1];
cx x[1],anc[0];
\end{lstlisting}
\caption{After inlining}
\end{subfigure}
\caption{Automatic ancilla sharing}
\label{fig:ancillamanagement}
\end{figure}

\subsection{Optimization}

Circuit optimization is necessary to produce efficient circuits which both utilize existing technology to the best of its ability, and to provide accurate resource estimates to guide the development of quantum algorithms and hardware. In contrast to other quantum computing software packages, \name was designed with circuit optimization as an integral part of the compiler. In this section we provide an overview of the optimization algorithms implemented: \texttt{simplify}, \texttt{rotation folding}, and \texttt{\edit{gray-synth}{CNOT resynthesis}}.

\paragraph{Gate simplifications}

The \texttt{simplify} optimization pass performs basic gate cancellations. In particular, it scans the program dependence graph and removes pairs of adjacent inverse gates whenever found, repeating until a fixpoint is reached. By (implicitly) using the program dependence graph rather than looking for gates adjacent in the AST, trivial commutations of gates acting on different qubits are modded out. As an example, \\
\centerline{\texttt{s x; h x; t y; h x; sdg x;}} \\
reduces to \texttt{t y;} with the \texttt{simplify} optimization pass. 

In general, because each pair of eliminated gates may open up other simplifications, the fixpoint computation may run $O(l)$ times, where $l$ is the number of gates in the program. In some cases this extra cost -- making the optimization quadratic in the length of the program -- may be prohibitive. In these cases, the user can opt to run single-pass simplifications instead of repeating until a fixpoint is reached.

\paragraph{Rotation folding}

The main optimization of the \name compiler is an extended implementation of Fang and Chen's $T$-count optimization\footnote{\edit{}{While $T$-count optimization itself is not particularly important in the context of NISQ programs, it has been shown \cite{aam18, nrscm17} that reducing the number of $T$ and other phase gates can frequently lead to significant CNOT reductions, which are much more costly in a NISQ setting.}} algorithm \cite{zc19}. Their algorithm reframes the problem of merging $T$ gates in Clifford+$T$ circuits in the \emph{Pauli sum} view of Gosset \emph{et al.} \cite{gkmr14}, in contrast to the \emph{phase polynomial} approach \cite{amm14, nrscm17}. We give a brief overview of their algorithm here.

Recall that the $T$ gate can be written as the following sum of Pauli gates:
\[
	T := \frac{1 + e^{i\pi/4}}{2}I + \frac{1 - e^{i\pi/4}}{2}Z.
\]
Following \cite{zc19} we write $R(P)$, where $P$ is an $n$-qubit Pauli operator possibly with a phase, for the Pauli sum 
\[
	R(P):=\frac{1 + e^{i\pi/4}}{2}I_n + \frac{1 - e^{i\pi/4}}{2}P
\]
As Clifford gates permute Pauli operators by conjugation, it can be observed that the following commutation rules hold for any Clifford gate $C$ and Paulis $P, P'$: 
\begin{align}
	CR(P) &= R(CPC^\dagger)C\label{eq:c1} \\
	PP'=P'P \implies R(P)R(P') &= R(P')R(P)\label{eq:c2}
\end{align}
Moreover, since the following equations hold:
\begin{align}
	R(P)R(P) &=R(P)^2\label{eq:e1} \\
	R(P)R(-P) &= I\label{eq:e2}
\end{align} 
$T$ gates can be merged by repeatedly applying commutations \ref{eq:c1} and \ref{eq:c2} and merging with any rotations satisfying \ref{eq:e1} or \ref{eq:e2}.

The \name implementation provides a number of extensions to Fang and Chen's algorithm -- notably, to handle $X$-, $Y$-, and $Z$-axis rotations \emph{of any angle} natively, and to allow arbitrary gates outside of the Clifford+$\{R_X, R_Y, R_Z\}$ gate set. This allows the optimization to be performed directly over arbitrary QASM programs, and further adds \edit{}{additional} applicability to NISQ-style circuits which use rotations of general angles in all Pauli axes. To extend Fang and Chen's algorithm in these ways we write $R(\theta, P)$ for a rotation of angle $\theta$ around the Pauli $P$, that is
\[
	R(\theta, P):=\frac{1 + e^{i\theta}}{2}I_n + \frac{1 - e^{i\theta}}{2}P,
\]
and then extend the commutation rules to
\begin{align}
	CR(\theta, P) &= R(\theta, CPC^\dagger)C\label{eq:cx1} \\
	PP'=P'P \implies R(\theta, P)R(\theta', P') &= R(\theta', P')R(\theta, P)\label{eq:cx2} \\
	\text{$U$ and $P$ act non-trivially on disjoint qubits} \implies UR(\theta, P) &= R(\theta, P)U\label{eq:cx3}
\end{align}
where $U$ refers to an arbitrary unitary gate.
The merging equations are likewise extended to
\begin{align}
	R(\theta, P)R(\theta', P) &=R(\theta + \theta', P)\label{eq:ce1} \\
	R(\theta, P)R(\theta', -P) &= e^{i\theta'}R(\theta - \theta', P)\label{eq:ce2}
\end{align}
which are both easy to verify.

Internally, \name provides a library for working with circuits in the Pauli sum representation, with classes for generic Clifford operators, Pauli rotations and uninterpreted gates, and methods for testing commutations and rotation merging. The \texttt{rotation folding} optimization is implemented as a Visitor on the AST which builds a representation of each basic block (i.e. gate bodies and the main program body) as a circuit in the Pauli sum representation and applies the above equations to determine which gates can be merged or cancelled. \edit{}{\Cref{fig:rotfol} shows an example of \name's rotation folding optimization. Our implementation can also optionally be configured to ignore irrelevant global phases.}

\begin{figure}
\centering
\begin{subfigure}[b]{0.49\textwidth}
\centering
\begin{lstlisting}
OPENQASM 2.0;
include "qelib1.inc";

qreg q[3];

t q[0];
t q[0];

rx(0.3) q[1];
h q[1];
rz(-0.2) q[1];
h q[1];
\end{lstlisting}
\caption{Before rotation folding}
\end{subfigure}
\begin{subfigure}[b]{0.49\textwidth}
\centering
\begin{lstlisting}
OPENQASM 2.0;
include "qelib1.inc";

qreg q[3];

s q[0];

h q[1];
rz(0.1) q[1];
h q[1];


\end{lstlisting}
\caption{After rotation folding}
\end{subfigure}
\caption{Folding rotation gates}
\label{fig:rotfol}
\end{figure}

\paragraph{\edit{CNOT-dihedral synthesis}{CNOT resynthesis}}

\edit{Currently implemented as a hardware mapping pass, the hardware-independent 
\texttt{gray-synth} optimization is accessible by mapping to a fully-connected device with
the \emph{Steiner tree mapping algorithm}, described later in Sec.~\ref{sec:mapping}. 
The \texttt{gray-synth} optimization attempts to reduce the number of~\texttt{cx} (i.e. 
CNOT) gates in the program by performing the CNOT-dihedral resynthesis algorithm of Amy,
Azimzadeh, and Mosca~\cite{aam18}}{In contrast to fault tolerant quantum computing, where 
small-angle $R_Z$ gates dominate the cost of quantum programs, in NISQ settings it is 
preferable to reduce the number of two-qubit gates, typically the CNOT gate. For these 
purposes \name includes an optimization pass -- \texttt{CNOT resynthesis} -- to reduce the
number of \texttt{cx} (i.e. CNOT) gates in a program by performing the CNOT-dihedral 
algorithm \texttt{Gray-synth} of Amy, Azimzadeh, and Mosca~\cite{aam18}}.
Specifically, 
the algorithm computes CNOT-dihedral blocks -- circuit blocks containing only CNOT, X, and
arbitrary phase rotations -- and then resynthesizes these blocks using using a Gray code 
inspired algorithm to construct an efficient tour of the necessary rotations. For more 
details on the algorithm, the reader is directed to \cite{aam18}.

The \texttt{\edit{gray-synth}{CNOT resynthesis}} algorithm is highly dependent on the number of $R_Z$ gates, and hence is typically most effective when used \emph{after} a \texttt{rotation folding} pass. In contrast to the implementation in \cite{aam18}, CNOT-dihedral blocks are computed greedily, and so not all foldable $R_Z$ are found with the \texttt{\edit{gray-synth}{CNOT resynthesis}} algorithm alone. \edit{A planned future extension will make this pass directly accessible as a high-level optimization pass, rather than as a mapping pass.}{While the \texttt{CNOT resynthesis} optimization can be used to optimize \texttt{cx} gates in a device-independent way, \name also includes an extension of \texttt{Gray-synth}, described in \Cref{sec:mapping}, which performs hardware mapping simultaneously with CNOT resynthesis to achieve better CNOT counts when the connectivity of the device is known.}

\subsection{Hardware mapping}\label{sec:mapping}

The NISQ era of quantum computing \cite{p18} carries with it specific hardware challenges -- notably, that of efficiently \emph{mapping} or \emph{routing} a quantum program onto a hardware device with constrained two-qubit interactions and noisy gates. In particular, this involves (1) mapping qubits of the program to physical qubits of the device, and (2) rewriting the circuit so that all two-qubit gates act on coupled qubits, and further satisfy the direction of the coupling in the case of directed topologies. \edit{}{While a complete overview of the range of existing hardware mapping techniques is beyond the scope of this work, \cite{cddkss19} provides a comprehensive review of existing permutation-based methods and performance comparisons.}

The \name compiler performs hardware mapping in two stages -- by first selecting an initial mapping from qubits of the program to physical qubits as in \cite{p19}, and then adjusting each two-qubit gate to conform to the given device topology. In this section we describe the algorithms for each stage implemented in \name. Currently, only physical CNOT gates are supported by \name.

\paragraph{Devices}

Devices in the \name toolchain are instances of the \texttt{Device} class, which at minimum specifies a number $n$ of qubits addressable on the device with addresses $0,\dots, n-1$, and a digraph where each directed edge represents an admissible CNOT gate with target at the edge's endpoint. A device may additionally specify average one- and two-qubit gate fidelities for each qubit and digraph edge, respectively, as floating-point numbers between $0$ and $1$. The \texttt{Device} class further contains a number of useful utilities for mapping circuits to devices with or without known fidelities; notably, the ability to retrieve the available couplings in order of decreasing fidelity, as well as fidelity-weighted shortest-path computation and approximation of minimal weighted Steiner trees -- trees spanning a subset of nodes.

While at present devices are hard-coded, built-in devices include the 8- and 16-qubit Rigetti Agave and Aspen4 chips, respectively, the 20-qubit IBM Tokyo device, a generic 9-qubit square lattice and fully-connected devices for any number of qubits.

\subsubsection{Initial layout generation}

It is known that the efficiency of hardware mapping is highly dependent on the chosen initial placement of qubits \cite{p19}. While better gate counts can often be achieved if the mapping algorithm is allowed to modify the initial placement \cite{zpw18, ldx19}, a good initial layout can reduce CNOT counts by over 50\% \cite{km19}.

\Name currently implements three layout generation algorithms: \texttt{linear}, \texttt{eager}, and \texttt{best-fit}. The \texttt{linear} layout generator functions as a basic layout, whereby physical qubits are assigned in-order as virtual qubits are allocated. By contrast, the \texttt{eager} and \texttt{best-fit} algorithms attempt to generate an initial layout which has a high degree of overlap between the CNOT gates present in the program, and the couplings present in the device.

The \texttt{eager} layout generator assigns highest-fidelity couplings on a first-come, first-serve basis. In particular, when a CNOT gate is encountered in the circuit, the highest-fidelity coupling which is \emph{compatible} with the control and target -- that is, doesn't invalidated previous assignments of the control or target to physical qubits -- is chosen. This strategy typically results in lower CNOT counts compared to the \texttt{linear} strategy when combined with basic swapping for CNOT mapping.

To generate a better initial layout for CNOT mapping algorithms which are \emph{not} based on local qubit swapping, an additional \texttt{best-fit} layout generation algorithm is implemented in \name. The \texttt{best-fit} algorithm, in contrast to the \texttt{linear} and \texttt{eager} strategies first scans the entire program before assigning physical qubits to virtual ones. In particular, it builds a histogram of couplings between virtual qubits, assigning the highest-fidelity couplings to virtual qubits with the most CNOT gates between them. Experimentally, we found that such an initial layout works best when qubits are not permuted intermittently by the CNOT mapping algorithm.

\begin{figure}
    \centering
    \begin{subfigure}[b]{0.30\textwidth}
        \centering
        \begin{lstlisting}
     OPENQASM 2.0;

     qreg q[9];
     CX q[2],q[1];
     CX q[2],q[1];
     CX q[6],q[8];
     CX q[7],q[3];
     CX q[7],q[3];
     CX q[7],q[3];
     CX q[5],q[7];
     CX q[5],q[7];
     CX q[5],q[7];
     CX q[5],q[7];
     CX q[4],q[0];
        \end{lstlisting}
        \caption{Initial circuit}
        \label{fig:mappinginitial}
    \end{subfigure}
    \begin{subfigure}[b]{0.50\textwidth}
        \centering
        \begin{tikzpicture}
            \node[circle, draw=black] (1) {$q_0$};
            \node[circle, draw=black, right=of 1] (2) {$q_1$};
            \node[circle, draw=black, right=of 2] (3) {$q_2$};
            \node[circle, draw=black, below=of 3] (4) {$q_3$};
            \node[circle, draw=black, left=of 4] (5) {$q_4$};
            \node[circle, draw=black, left=of 5] (6) {$q_5$};
            \node[circle, draw=black, below=of 6] (7) {$q_6$};
            \node[circle, draw=black, right=of 7] (8) {$q_7$};
            \node[circle, draw=black, right=of 8] (9) {$q_8$};
            \draw[thick] (1) -- node[above] {0.93} (2);
            \draw[thick] (1) -- node[left] {0.75} (6);
            \draw[thick] (2) -- node[above] {0.75} (3);
            \draw[thick] (2) -- node[left] {0.91} (5);
            \draw[thick] (3) -- node[left] {0.74} (4);
            \draw[thick] (4) -- node[above] {0.76} (5);
            \draw[thick] (4) -- node[left] {0.72} (9);
            \draw[thick] (5) -- node[above] {0.77} (6);
            \draw[thick] (5) -- node[left] {0.90} (8);
            \draw[thick] (6) -- node[left] {0.76} (7);
            \draw[thick] (7) -- node[above] {0.89} (8);
            \draw[thick] (8) -- node[above] {0.87} (9);
        \end{tikzpicture}
        \vspace{1em}
        \caption{Device topology \& fidelities}
        \label{fig:mappingdevice}
    \end{subfigure}
    
    \begin{subfigure}[b]{0.32\textwidth}
        \centering
    \begin{tikzpicture}[cell/.style={rectangle,draw=black,minimum width=2em}]
    \matrix(in)[matrix of nodes,column 1/.style={nodes={cell}}]
    {0 \\ 1 \\ 2 \\ 3 \\ 4 \\ 5 \\ 6 \\ 7 \\ 8 \\};
    \matrix(out)[right=of in, matrix of nodes,column 1/.style={nodes={cell}}]
    {0 \\ 1 \\ 2 \\ 3 \\ 4 \\ 5 \\ 6 \\ 7 \\ 8 \\};
    \node[above=-.5em of in] {$q_{\text{in}}$};
    \node[above=-.5em of out] {$q_{\text{out}}$};
    \draw[->, thick, draw=black] (in-1-1) -- (out-1-1);
    \draw[->, thick, draw=black] (in-2-1) -- (out-2-1);
    \draw[->, thick, draw=black] (in-3-1) -- (out-3-1);
    \draw[->, thick, draw=black] (in-4-1) -- (out-4-1);
    \draw[->, thick, draw=black] (in-5-1) -- (out-5-1);
    \draw[->, thick, draw=black] (in-6-1) -- (out-6-1);
    \draw[->, thick, draw=black] (in-7-1) -- (out-7-1);
    \draw[->, thick, draw=black] (in-8-1) -- (out-8-1);
    \draw[->, thick, draw=black] (in-9-1) -- (out-9-1);
    \end{tikzpicture}
        \caption{Linear layout}
        \label{fig:mappinglinear1}
    \end{subfigure}
    \begin{subfigure}[b]{0.32\textwidth}
        \centering
    \begin{tikzpicture}[cell/.style={rectangle,draw=black,minimum width=2em}]
    \matrix(in)[matrix of nodes,column 1/.style={nodes={cell}}]
    {0 \\ 1 \\ 2 \\ 3 \\ 4 \\ 5 \\ 6 \\ 7 \\ 8 \\};
    \matrix(out)[right=of in, matrix of nodes,column 1/.style={nodes={cell}}]
    {0 \\ 1 \\ 2 \\ 3 \\ 4 \\ 5 \\ 6 \\ 7 \\ 8 \\};
    \node[above=-.5em of in] {$q_{\text{in}}$};
    \node[above=-.5em of out] {$q_{\text{out}}$};
    \draw[->, thick, draw=black] (in-1-1.east) -- (out-9-1.west);
    \draw[->, thick, draw=black] (in-2-1.east) -- (out-2-1.west);
    \draw[->, thick, draw=black] (in-3-1.east) -- (out-1-1.west);
    \draw[->, thick, draw=black] (in-4-1.east) -- (out-7-1.west);
    \draw[->, thick, draw=black] (in-5-1.east) -- (out-4-1.west);
    \draw[->, thick, draw=black] (in-6-1.east) -- (out-3-1.west);
    \draw[->, thick, draw=black] (in-7-1.east) -- (out-5-1.west);
    \draw[->, thick, draw=black] (in-8-1.east) -- (out-6-1.west);
    \draw[->, thick, draw=black] (in-9-1.east) -- (out-8-1.west);
    \end{tikzpicture}
        \caption{Eager layout}
        \label{fig:mappingeager}
    \end{subfigure}
    \begin{subfigure}[b]{0.32\textwidth}
        \centering
    \begin{tikzpicture}[cell/.style={rectangle,draw=black,minimum width=2em}]
    \matrix(in)[matrix of nodes,column 1/.style={nodes={cell}}]
    {0 \\ 1 \\ 2 \\ 3 \\ 4 \\ 5 \\ 6 \\ 7 \\ 8 \\};
    \matrix(out)[right=of in, matrix of nodes,column 1/.style={nodes={cell}}]
    {0 \\ 1 \\ 2 \\ 3 \\ 4 \\ 5 \\ 6 \\ 7 \\ 8 \\};
    \node[above=-.5em of in] {$q_{\text{in}}$};
    \node[above=-.5em of out] {$q_{\text{out}}$};
    \draw[->, thick, draw=black] (in-1-1.east) -- (out-9-1.west);
    \draw[->, thick, draw=black] (in-2-1.east) -- (out-8-1.west);
    \draw[->, thick, draw=black] (in-3-1.east) -- (out-7-1.west);
    \draw[->, thick, draw=black] (in-4-1.east) -- (out-5-1.west);
    \draw[->, thick, draw=black] (in-5-1.east) -- (out-6-1.west);
    \draw[->, thick, draw=black] (in-6-1.east) -- (out-1-1.west);
    \draw[->, thick, draw=black] (in-7-1.east) -- (out-3-1.west);
    \draw[->, thick, draw=black] (in-8-1.east) -- (out-2-1.west);
    \draw[->, thick, draw=black] (in-9-1.east) -- (out-4-1.west);
    \end{tikzpicture}
        \caption{Best-fit layout}
        \label{fig:mappinglinear2}
    \end{subfigure}
    \caption{Laying out a circuit on a $9$-qubit square lattice}
    \label{fig:mapping}
\end{figure}
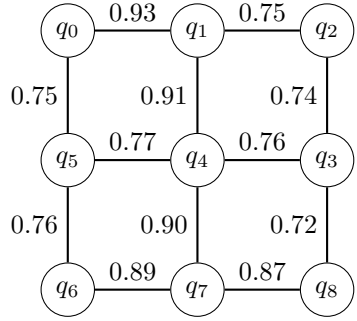
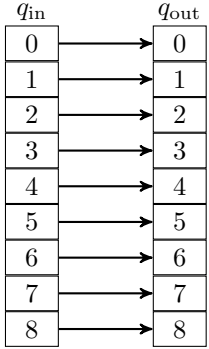
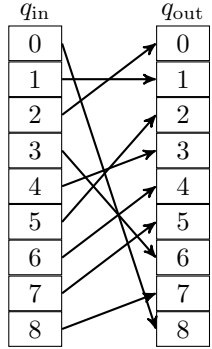
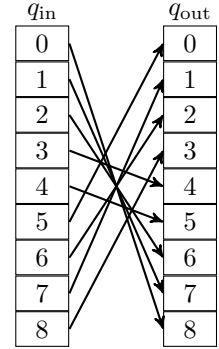

To illustrate the different initial placement approaches, \Cref{fig:mapping} gives an example of each layout generation algorithm applied to a circuit for a simple square lattice shown in \Cref{fig:mappingdevice}.

\subsubsection{CNOT mapping}

\begin{figure}
    \centering
    \[\Qcircuit @C=1em @R=.4em @!R {
 		& \qw & \ctrl{3} & \qw & \qw \\
 		& \qw & \qw & \qw & \qw \\
 		& \qw & \qw & \qw & \qw \\
 		& \qw & \targ & \qw & \qw
	}
	\quad
	\raisebox{-1.8em}{$\rightarrow$}
	\quad
	\Qcircuit @C=1em @R=.4em @!R {
 		& \qw & \qswap & \qw & \qw & \qw & \qswap & \qw & \qw \\
 		& \qw & \qw \qwx & \qswap & \ctrl{2} & \qswap & \qw \qwx & \qw & \qw \\
 		& \qw & \qswap \qwx & \qswap \qwx & \qw & \qswap \qwx & \qswap \qwx & \qw & \qw \\
 		& \qw & \qw & \qw & \targ & \qw & \qw & \qw & \qw
	}
	\quad
	\raisebox{-1.8em}{$\rightarrow$}
	\quad
	\Qcircuit @C=1em @R=.4em @!R {
 		& \qw & \ctrl{2} & \targ & \ctrl{2} & \qw & \qw & \qw & \qw & \qw & \qw & \qw & \ctrl{2} & \targ & \ctrl{2} & \qw & \qw \\
 		& \qw & \qw & \qw & \qw & \ctrl{1} & \targ & \ctrl{1} & \ctrl{2} & \ctrl{1} & \targ & \ctrl{1} & \qw & \qw & \qw & \qw & \qw \\
 		& \qw & \targ & \ctrl{-2} & \targ & \targ & \ctrl{-1} & \targ & \qw & \targ & \ctrl{-1} & \targ & \targ & \ctrl{-2} & \targ & \qw & \qw \\
 		& \qw & \qw & \qw & \qw & \qw & \qw & \qw & \targ & \qw & \qw & \qw & \qw \qw & \qw & \qw & \qw & \qw
	}\]
    \caption{Mapping a CNOT gate via local swaps to a device with couplings $(0,2), (1,2), (1,3)$}
    \label{fig:swapping}
\end{figure}
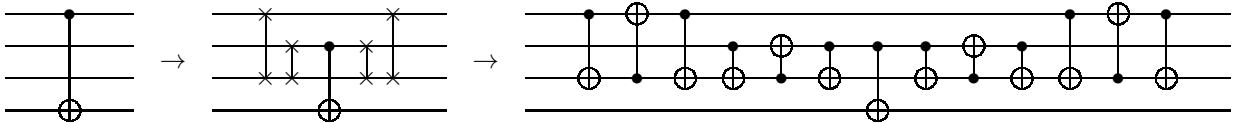

The problem of mapping two-qubit gates to a topologically constrained architecture has received a great deal of attention recently \cite{km19,ldx19,ngm19,p19,zpw18,cddkss19}. Most common techniques (e.g., the IBM-QX contest-winning \cite{zpw18}) rely on inserting swap gates, or more generally permutations, so that the a given two-qubit gate or set of gates satisfies the device topology. \Cref{fig:swapping} shows an example of this technique.

\Name implements a version of permutation-based mapping (\texttt{swap}) where for a CNOT gate between uncoupled qubits, the endpoints are swapped along the shortest (weighted) path in the coupling graph until they are adjacent. In the case of directed edges, Hadamard gates are inserted to flip the control and target of a CNOT. Rather than swap the qubits back to their original position as in \Cref{fig:swapping}, the resulting permutation is propagated through the rest of the circuit.

\paragraph{Steiner tree mapping}

An alternative to permutation-based mappings which has recently been gaining popularity is \emph{topologically-constrained synthesis} \cite{km19}. With this technique, a circuit or subcircuit is re-synthesized using circuit synthesis techniques that \emph{directly} account for the topology of the intended architecture. For circuits consisting solely of CNOT gates, \cite{km19} and \cite{ngm19} simultaneously developed methods of synthesizing efficient circuits satisfying a given topology by performing \emph{constrained Gaussian elimination}, whereby the rows which can be added to one another, corresponding to qubits, are restricted by the underlying architecture. These results show that in the case of CNOT -- or linear reversible circuits -- constrained Gaussian elimination typically results in lower CNOT counts than permutation-based techniques \cite{km19}. Both results further sketch extensions to the topologically-constrained synthesis of CNOT-dihedral circuits. Along with the \texttt{swap} mapping algorithm, \name includes a mapping algorithm (\texttt{steiner}) based on constrained CNOT and CNOT-dihedral synthesis in the style of \cite{km19} and \cite{ngm19}. We give a brief overview of the \texttt{steiner} mapper here.

\begin{figure}
\begin{subfigure}{0.49\textwidth}
    
    \[\hspace{-3em}\Qcircuit @C=1em @R=.4em @!R {
 		\lstick{x_1} & \ctrl{2} & \qw & \qw & \targ & \qw & \qw & \rstick{\!\!\!\!x_1} \\
 		\lstick{x_2} & \qw & \targ & \qw & \qw & \ctrl{1} & \qw & \rstick{\!\!\!\!x_2\oplus x_4} \\
 		\lstick{x_3} & \targ & \qw & \ctrl{1} & \ctrl{-2} & \targ & \qw & \rstick{\!\!\!\!x_1\oplus x_2\oplus x_3} \\
 		\lstick{x_4} & \qw & \ctrl{-2} & \targ & \qw & \qw & \qw & \rstick{\!\!\!\!x_1\oplus x_3\oplus x_4}
	}\]
\caption{A circuit over CNOT gates}
\label{fig:linrevcnot}
\end{subfigure}
\begin{subfigure}{0.49\textwidth}
\[
    \left[
    \begin{array}{cccc}
        1 & 0 & 0 & 0 \\
        0 & 1 & 0 & 1 \\
        1 & 1 & 1 & 0 \\
        1 & 0 & 1 & 1
    \end{array}
    \right]
    \left[
    \begin{array}{c}
        x_1 \\
        x_2 \\
        x_3 \\
        x_4
    \end{array}
    \right]
    =
    \left[
    \begin{array}{l}
        x_1 \\
        x_2 \oplus x_4 \\
        x_1 \oplus x_2 \oplus x_3 \\
        x_1 \oplus x_3 \oplus x_4
    \end{array}
    \right]
\]
\caption{The corresponding binary matrix}
\label{fig:linrevmatrix}
\end{subfigure}
\caption{Linear reversible circuits}
\label{fig:linrev}
\end{figure}
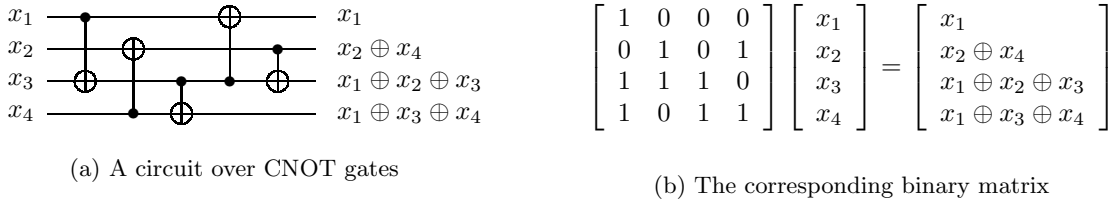

The standard method\footnote{With a small modification, this method is in fact asymptotically optimal \cite{pmh08}.} of synthesizing an $n$-qubit CNOT circuit, is to perform Gaussian elimination on the $n\times n$ binary matrix giving the classical function (see \Cref{fig:linrev}) and reverse the row operations, corresponding to CNOT gates. When the hardware topology is constrained however, it may not be possible to ``zero-out'' all the off-diagonal entries of a column by adding the pivot row to them directly. Instead, a path in the coupling graph from the pivot to each row with a leading $1$ may be used by first \emph{filling} in $1$'s along the path by applying CNOT gates, then \emph{flushing} by applying CNOT gates along the path in reverse. For example, with the ``straight-line'' topology, the $1$ in entry $(2,0)$ of the matrix in \Cref{fig:linrevmatrix} can be eliminated by first filling in $1$'s along the shortest path from $0$ to $2$, then removing them in reverse:
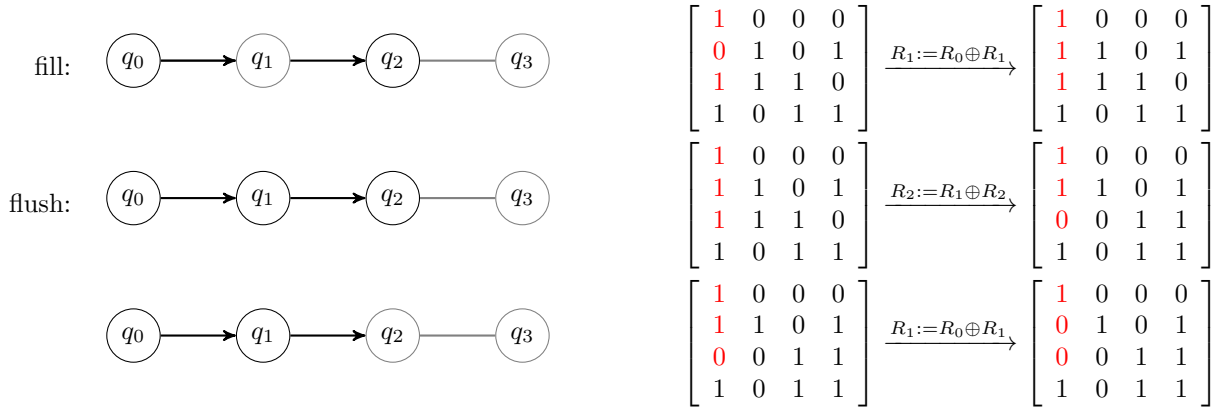
\begin{align*}
    \text{fill:}\quad
    &
    \raisebox{-0.5em}{
    \begin{tikzpicture}
        \node[circle, draw=black] (1) {$q_0$};
        \node[circle, draw=gray, right=of 1] (2) {$q_1$};
        \node[circle, draw=black, right=of 2] (3) {$q_2$};
        \node[circle, draw=gray, right=of 3] (4) {$q_3$};
        \draw[thick, draw=gray] (1) -- (2);
        \draw[->, thick, draw=black] (1) -- (2);
        \draw[->, thick, draw=black] (2) -- (3);
        \draw[thick, draw=gray] (3) -- (4);
    \end{tikzpicture}}
    & \qquad\qquad
	\left[
	\begin{array}{cccc}
        \color{red}1 & 0 & 0 & 0 \\
        \color{red}0 & 1 & 0 & 1 \\
        \color{red}1 & 1 & 1 & 0 \\
        1 & 0 & 1 & 1
    \end{array}
    \right]
    \xrightarrow{R_1:=R_0\oplus R_1}
    \left[
    \begin{array}{cccc}
        \color{red}1 & 0 & 0 & 0 \\
        \color{red}1 & 1 & 0 & 1 \\
        \color{red}1 & 1 & 1 & 0 \\
        1 & 0 & 1 & 1
    \end{array}
    \right] \\
    \text{flush:}\quad &
    \raisebox{-0.5em}{
    \begin{tikzpicture}
        \node[circle, draw=black] (1) {$q_0$};
        \node[circle, draw=black, right=of 1] (2) {$q_1$};
        \node[circle, draw=black, right=of 2] (3) {$q_2$};
        \node[circle, draw=gray, right=of 3] (4) {$q_3$};
        \draw[->, thick, draw=black] (1) -- (2);
        \draw[->, thick, draw=black] (2) -- (3);
        \draw[thick, draw=gray] (3) -- (4);
    \end{tikzpicture}}
    & \qquad\qquad
	\left[
	\begin{array}{cccc}
        \color{red}1 & 0 & 0 & 0 \\
        \color{red}1 & 1 & 0 & 1 \\
        \color{red}1 & 1 & 1 & 0 \\
        1 & 0 & 1 & 1
    \end{array}
    \right]
    \xrightarrow{R_2:=R_1\oplus R_2}
    \left[
    \begin{array}{cccc}
        \color{red}1 & 0 & 0 & 0 \\
        \color{red}1 & 1 & 0 & 1 \\
        \color{red}0 & 0 & 1 & 1 \\
        1 & 0 & 1 & 1
    \end{array}
    \right] \\
    &
    \raisebox{-0.5em}{
    \begin{tikzpicture}
        \node[circle, draw=black] (1) {$q_0$};
        \node[circle, draw=black, right=of 1] (2) {$q_1$};
        \node[circle, draw=gray, right=of 2] (3) {$q_2$};
        \node[circle, draw=gray, right=of 3] (4) {$q_3$};
        \draw[->, thick, draw=black] (1) -- (2);
        \draw[->, thick, draw=black] (2) -- (3);
        \draw[thick, draw=gray] (3) -- (4);
    \end{tikzpicture}}
    & \qquad\qquad
	\left[
	\begin{array}{cccc}
        \color{red}1 & 0 & 0 & 0 \\
        \color{red}1 & 1 & 0 & 1 \\
        \color{red}0 & 0 & 1 & 1 \\
        1 & 0 & 1 & 1
    \end{array}
    \right]
    \xrightarrow{R_1:=R_0\oplus R_1}
    \left[
    \begin{array}{cccc}
        \color{red}1 & 0 & 0 & 0 \\
        \color{red}0 & 1 & 0 & 1 \\
        \color{red}0 & 0 & 1 & 1 \\
        1 & 0 & 1 & 1
    \end{array}
    \right]
\end{align*}

To zero-out all the leading $1$'s, excluding the pivot, a \emph{tree} with root at the pivot qubit and endpoints at all rows with a leading $1$ can be used instead. As noted in \cite{km19, ngm19}, computing a minimal such tree is the well-known \emph{Steiner tree problem}, which is NP-hard in general but admits effective polynomial-time approximations, notably via all-pairs-shortest-paths and minimal spanning tree algorithms. Zeroing all non-pivot rows for a given column proceeds similarly, by first filling $1$'s into every node of the tree by adding rows along every edge leading to a $0$ -- called \emph{Steiner points} -- then zero-ing all non-root nodes by traversing the tree and adding rows along edges in reverse. An example is given below with a minimal tree spanning $\{q_0, q_2, q_4\}$ with edges shown in bold:
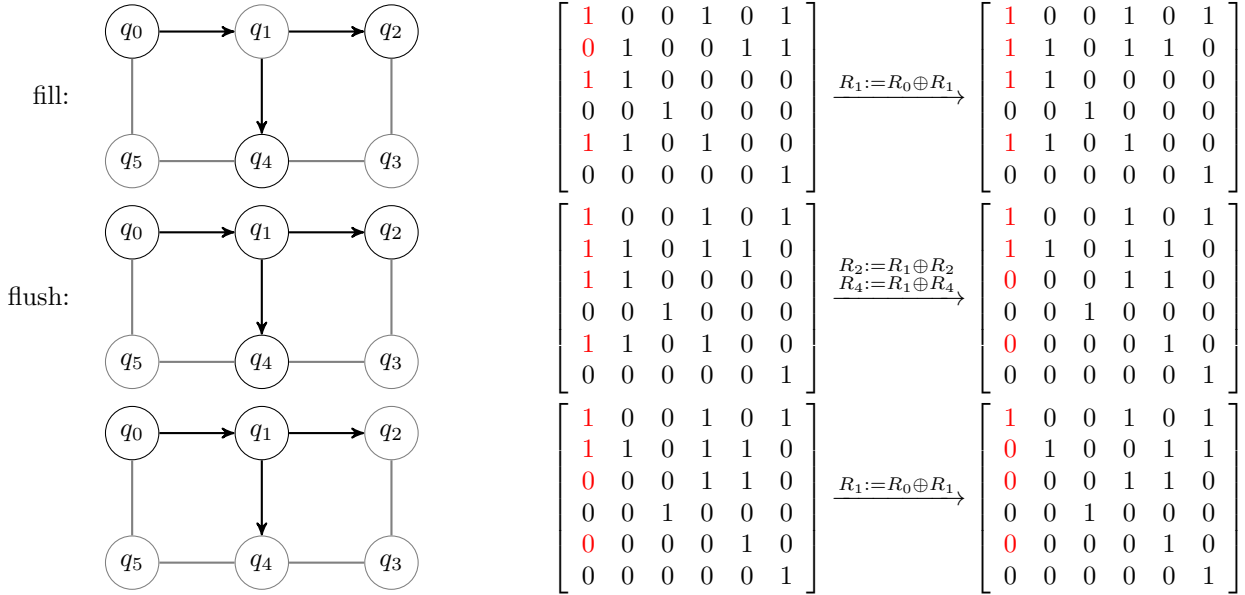
\begin{align*}
    \text{fill:}\quad
    &
    \raisebox{-3.2em}{
    \begin{tikzpicture}
        \node[circle, draw=black] (1) {$q_0$};
        \node[circle, draw=gray, right=of 1] (2) {$q_1$};
        \node[circle, draw=black, right=of 2] (3) {$q_2$};
        \node[circle, draw=gray, below=of 3] (4) {$q_3$};
        \node[circle, draw=black, left=of 4] (5) {$q_4$};
        \node[circle, draw=gray, left=of 5] (6) {$q_5$};
        \draw[->, thick, draw=black] (1) -- (2);
        \draw[thick, draw=gray] (1) -- (6);
        \draw[->, thick, draw=black] (2) -- (3);
        \draw[->, thick, draw=black] (2) -- (5);
        \draw[thick, draw=gray] (3) -- (4);
        \draw[thick, draw=gray] (4) -- (5);
        \draw[thick, draw=gray] (5) -- (6);
    \end{tikzpicture}}
    & \qquad\qquad
	\left[
	\begin{array}{cccccc}
        \color{red} 1 & 0 & 0 & 1 & 0 & 1 \\
        \color{red} 0 & 1 & 0 & 0 & 1 & 1 \\
        \color{red} 1 & 1 & 0 & 0 & 0 & 0 \\
        0 & 0 & 1 & 0 & 0 & 0 \\
        \color{red} 1 & 1 & 0 & 1 & 0 & 0 \\
        0 & 0 & 0 & 0 & 0 & 1
    \end{array}
    \right]
    \xrightarrow{R_1:=R_0\oplus R_1}
    \left[
	\begin{array}{cccccc}
        \color{red} 1 & 0 & 0 & 1 & 0 & 1 \\
        \color{red} 1 & 1 & 0 & 1 & 1 & 0 \\
        \color{red} 1 & 1 & 0 & 0 & 0 & 0 \\
        0 & 0 & 1 & 0 & 0 & 0 \\
        \color{red} 1 & 1 & 0 & 1 & 0 & 0 \\
        0 & 0 & 0 & 0 & 0 & 1
    \end{array}
    \right] \\
    \text{flush:}\quad
    &
    \raisebox{-3.2em}{
    \begin{tikzpicture}
        \node[circle, draw=black] (1) {$q_0$};
        \node[circle, draw=black, right=of 1] (2) {$q_1$};
        \node[circle, draw=black, right=of 2] (3) {$q_2$};
        \node[circle, draw=gray, below=of 3] (4) {$q_3$};
        \node[circle, draw=black, left=of 4] (5) {$q_4$};
        \node[circle, draw=gray, left=of 5] (6) {$q_5$};
        \draw[->, thick, draw=black] (1) -- (2);
        \draw[thick, draw=gray] (1) -- (6);
        \draw[->, thick, draw=black] (2) -- (3);
        \draw[->, thick, draw=black] (2) -- (5);
        \draw[thick, draw=gray] (3) -- (4);
        \draw[thick, draw=gray] (4) -- (5);
        \draw[thick, draw=gray] (5) -- (6);
    \end{tikzpicture}}
    & \qquad\qquad
	\left[
	\begin{array}{cccccc}
        \color{red} 1 & 0 & 0 & 1 & 0 & 1 \\
        \color{red} 1 & 1 & 0 & 1 & 1 & 0 \\
        \color{red} 1 & 1 & 0 & 0 & 0 & 0 \\
        0 & 0 & 1 & 0 & 0 & 0 \\
        \color{red} 1 & 1 & 0 & 1 & 0 & 0 \\
        0 & 0 & 0 & 0 & 0 & 1
    \end{array}
    \right]
    \xrightarrow{\substack{R_2:=R_1\oplus R_2 \\ R_4:=R_1 \oplus R_4}}
    \left[
	\begin{array}{cccccc}
        \color{red} 1 & 0 & 0 & 1 & 0 & 1 \\
        \color{red} 1 & 1 & 0 & 1 & 1 & 0 \\
        \color{red} 0 & 0 & 0 & 1 & 1 & 0 \\
        0 & 0 & 1 & 0 & 0 & 0 \\
        \color{red} 0 & 0 & 0 & 0 & 1 & 0 \\
        0 & 0 & 0 & 0 & 0 & 1
    \end{array}
    \right] \\
    &
    \raisebox{-3.2em}{
    \begin{tikzpicture}
        \node[circle, draw=black] (1) {$q_0$};
        \node[circle, draw=black, right=of 1] (2) {$q_1$};
        \node[circle, draw=gray, right=of 2] (3) {$q_2$};
        \node[circle, draw=gray, below=of 3] (4) {$q_3$};
        \node[circle, draw=gray, left=of 4] (5) {$q_4$};
        \node[circle, draw=gray, left=of 5] (6) {$q_5$};
        \draw[->, thick, draw=black] (1) -- (2);
        \draw[thick, draw=gray] (1) -- (6);
        \draw[->, thick, draw=black] (2) -- (3);
        \draw[->, thick, draw=black] (2) -- (5);
        \draw[thick, draw=gray] (3) -- (4);
        \draw[thick, draw=gray] (4) -- (5);
        \draw[thick, draw=gray] (5) -- (6);
    \end{tikzpicture}}
    & \qquad\qquad
	\left[
	\begin{array}{cccccc}
        \color{red} 1 & 0 & 0 & 1 & 0 & 1 \\
        \color{red} 1 & 1 & 0 & 1 & 1 & 0 \\
        \color{red} 0 & 0 & 0 & 1 & 1 & 0 \\
        0 & 0 & 1 & 0 & 0 & 0 \\
        \color{red} 0 & 0 & 0 & 0 & 1 & 0 \\
        0 & 0 & 0 & 0 & 0 & 1
    \end{array}
    \right]
    \xrightarrow{R_1:=R_0\oplus R_1}
    \left[
	\begin{array}{cccccc}
        \color{red} 1 & 0 & 0 & 1 & 0 & 1 \\
        \color{red} 0 & 1 & 0 & 0 & 1 & 1 \\
        \color{red} 0 & 0 & 0 & 1 & 1 & 0 \\
        0 & 0 & 1 & 0 & 0 & 0 \\
        \color{red} 0 & 0 & 0 & 0 & 1 & 0 \\
        0 & 0 & 0 & 0 & 0 & 1
    \end{array}
    \right]
\end{align*}

\paragraph{Crossing the diagonal}
The ``fill-then-flush'' method may fail when the computed tree contains nodes \emph{above} the diagonal, as this may propagate $1$'s to the left of the current column, as in the following example:
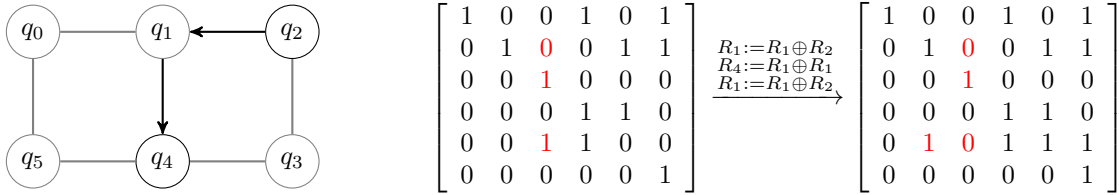
\[
    \raisebox{-3.2em}{
    \begin{tikzpicture}
        \node[circle, draw=gray] (1) {$q_0$};
        \node[circle, draw=gray, right=of 1] (2) {$q_1$};
        \node[circle, draw=black, right=of 2] (3) {$q_2$};
        \node[circle, draw=gray, below=of 3] (4) {$q_3$};
        \node[circle, draw=black, left=of 4] (5) {$q_4$};
        \node[circle, draw=gray, left=of 5] (6) {$q_5$};
        \draw[thick, draw=gray] (1) -- (2);
        \draw[thick, draw=gray] (1) -- (6);
        \draw[->, thick, draw=black] (3) -- (2);
        \draw[->, thick, draw=black] (2) -- (5);
        \draw[thick, draw=gray] (3) -- (4);
        \draw[thick, draw=gray] (4) -- (5);
        \draw[thick, draw=gray] (5) -- (6);
    \end{tikzpicture}}
    \qquad\qquad
	\left[
	\begin{array}{cccccc}
        1 & 0 & 0 & 1 & 0 & 1 \\
        0 & 1 & \color{red}0 & 0 & 1 & 1 \\
        0 & 0 & \color{red}1 & 0 & 0 & 0 \\
        0 & 0 & 0 & 1 & 1 & 0 \\
        0 & 0 & \color{red}1 & 1 & 0 & 0 \\
        0 & 0 & 0 & 0 & 0 & 1
    \end{array}
    \right]
    \xrightarrow{\substack{R_1:=R_1\oplus R_2 \\ R_4:=R_1 \oplus R_1 \\ R_1:=R_1\oplus R_2}}
    \left[
	\begin{array}{cccccc}
        1 & 0 & 0 & 1 & 0 & 1 \\
        0 & 1 & \color{red}0 & 0 & 1 & 1 \\
        0 & 0 & \color{red}1 & 0 & 0 & 0 \\
        0 & 0 & 0 & 1 & 1 & 0 \\
        0 & \color{red}1 & \color{red}0 & 1 & 1 & 1 \\
        0 & 0 & 0 & 0 & 0 & 1
    \end{array}
    \right]
\]
In \cite{km19}, the above-diagonal dependencies are handled by ordering the rows according to a \emph{Hamiltonian path} in the graph (if it exists), so that the matrix can be reduced to echelon form without crossing the diagonal. As not all possible topologies admit a Hamiltonian path -- and in general computing one is an NP-hard problem -- they also give a recursive method which works for arbitrary graphs. In contrast, \cite{ngm19} doesn't assume the existence of a Hamiltonian path and instead uses an \emph{uncompute} stage to effectively ``undo'' all changes to other columns.

The implementation in \name follows the method of \cite{ngm19}, with the exception that \emph{only changes to rows with (transitive) dependencies on above-diagonal rows are uncomputed}. In practice this reduces the number of CNOT gates required, as not every iteration will cross the diagonal.

\paragraph{Constrained Gray-synth}

The \texttt{Steiner} mapping algorithm in \name actually implements a more general form of re-synthesis, targeting CNOT-dihedral\footnote{Circuits over $\{\text{CNOT}, X, R_Z(\theta)\}$} circuits by using the \texttt{Gray-synth} CNOT-optimization algorithm \cite{aam18} extended with constrained Gaussian synthesis. Similar extensions were considered in \cite{km19} and \cite{ngm19} -- by comparison \name contains a full-scale implementation which operates on arbitrary circuits by generating \emph{synthesis events} whenever a non-CNOT-dihedral gate is encountered. In the case when no $R_Z(\theta)$ gates are present, the algorithm coincides with the basic constrained Gaussian synthesis.

Again, the implementation of constrained \texttt{Gray-synth} differs from those sketched in \cite{km19, ngm19}, so we give a brief overview of our method here. The \texttt{Gray-synth} algorithm functions by ordering a given set of \emph{parities} -- corresponding to the states ``being rotated on'' by phase gates -- of the inputs so that an efficient tour can be constructed. Moreover, this ordering is elaborated as the circuit is synthesized, by recursively partitioning and synthesizing the remaining parities and updating the remaining parities as the state is modified by the synthesized CNOT gates.

\Name performs constrained Gray-synthesis by delaying CNOT gates \emph{until a partition of size $1$ is reached}. Such a partition corresponds to the computation of a parity $x_i \mapsto x_i \oplus f(x_1,\dots, x_n)$ where $f$ is a linear (parity) function over $\{x_{j \neq i}\}$. Again, a (Steiner) tree rooted at $x_i$ and spanning $S=\{x_j \mid x_j \text{ appears in } f \}$ can be used to synthesize the above parity -- however, in contrast to the Gaussian elimination situation, where the root is added to each leaf, \emph{each leaf needs to be added to the root}. This is done by first adding each Steiner point (nodes in the tree but not in $S$) to its predecessors in breadth-first order, then adding each node to its predecessors in reverse breadth-first order. An example of this process is given below for the parity $x_0 \oplus x_1\oplus x_2\oplus x_4\oplus x_8$ rooted at $x_0$ on a square lattice. Note that the matrix in this case gives the function computed by the series of row additions (i.e. CNOT gates).
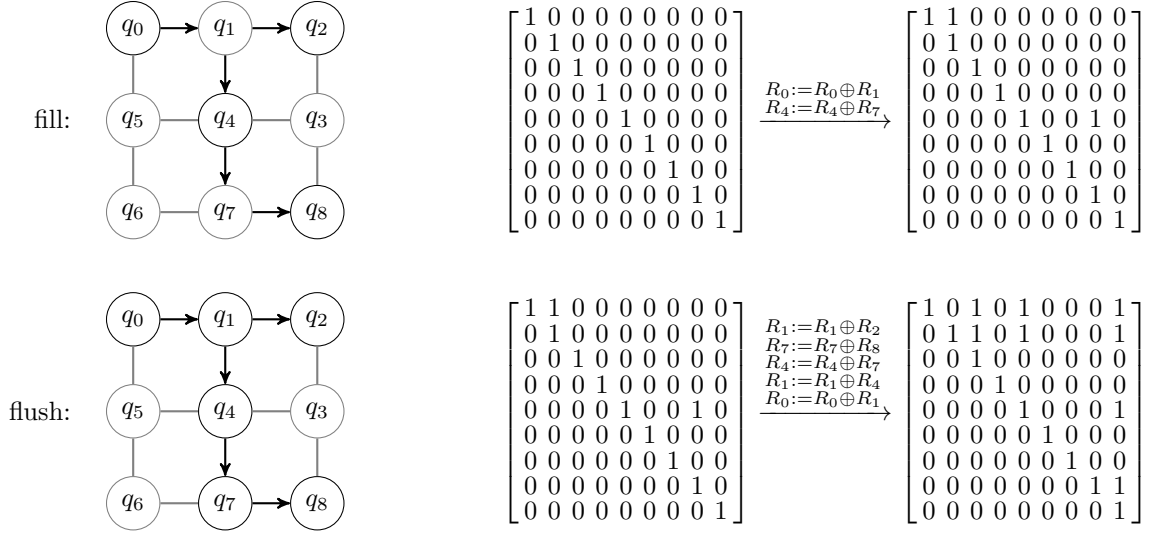
\begingroup 
\setlength\arraycolsep{2pt}
\def\arraystretch{.8}
\begin{align*}
    \text{fill:} \quad &
    \raisebox{-4.2em}{
    \begin{tikzpicture}[node distance = .5cm]
        \node[circle, draw=black] (1) {$q_0$};
        \node[circle, draw=gray, right=of 1] (2) {$q_1$};
        \node[circle, draw=black, right=of 2] (3) {$q_2$};
        \node[circle, draw=gray, below=of 3] (4) {$q_3$};
        \node[circle, draw=black, left=of 4] (5) {$q_4$};
        \node[circle, draw=gray, left=of 5] (6) {$q_5$};
        \node[circle, draw=gray, below=of 6] (7) {$q_6$};
        \node[circle, draw=gray, right=of 7] (8) {$q_7$};
        \node[circle, draw=black, right=of 8] (9) {$q_8$};
        \draw[->, thick, draw=black] (1) -- (2);
        \draw[thick, draw=gray] (1) -- (6);
        \draw[->, thick, draw=black] (2) -- (3);
        \draw[->, thick, draw=black] (2) -- (5);
        \draw[thick, draw=gray] (3) -- (4);
        \draw[thick, draw=gray] (4) -- (5);
        \draw[thick, draw=gray] (4) -- (9);
        \draw[thick, draw=gray] (5) -- (6);
        \draw[->, thick, draw=black] (5) -- (8);
        \draw[thick, draw=gray] (6) -- (7);
        \draw[thick, draw=gray] (7) -- (8);
        \draw[->, thick, draw=black] (8) -- (9);
    \end{tikzpicture}}
    & \qquad\qquad
	\left[
	\begin{array}{ccccccccc}
        1 & 0 & 0 & 0 & 0 & 0 & 0 & 0 & 0 \\
        0 & 1 & 0 & 0 & 0 & 0 & 0 & 0 & 0 \\
        0 & 0 & 1 & 0 & 0 & 0 & 0 & 0 & 0 \\
        0 & 0 & 0 & 1 & 0 & 0 & 0 & 0 & 0 \\
        0 & 0 & 0 & 0 & 1 & 0 & 0 & 0 & 0 \\
        0 & 0 & 0 & 0 & 0 & 1 & 0 & 0 & 0 \\
        0 & 0 & 0 & 0 & 0 & 0 & 1 & 0 & 0 \\
        0 & 0 & 0 & 0 & 0 & 0 & 0 & 1 & 0 \\
        0 & 0 & 0 & 0 & 0 & 0 & 0 & 0 & 1 \\
    \end{array}
    \right]
    \xrightarrow{\substack{R_0:=R_0\oplus R_1 \\ R_4:=R_4 \oplus R_7}}
    \left[
	\begin{array}{ccccccccc}
        1 & 1 & 0 & 0 & 0 & 0 & 0 & 0 & 0 \\
        0 & 1 & 0 & 0 & 0 & 0 & 0 & 0 & 0 \\
        0 & 0 & 1 & 0 & 0 & 0 & 0 & 0 & 0 \\
        0 & 0 & 0 & 1 & 0 & 0 & 0 & 0 & 0 \\
        0 & 0 & 0 & 0 & 1 & 0 & 0 & 1 & 0 \\
        0 & 0 & 0 & 0 & 0 & 1 & 0 & 0 & 0 \\
        0 & 0 & 0 & 0 & 0 & 0 & 1 & 0 & 0 \\
        0 & 0 & 0 & 0 & 0 & 0 & 0 & 1 & 0 \\
        0 & 0 & 0 & 0 & 0 & 0 & 0 & 0 & 1 \\
    \end{array}
    \right] \\ \\
    \text{flush:} \quad &
    \raisebox{-4.2em}{
    \begin{tikzpicture}[node distance = .5cm]
        \node[circle, draw=black] (1) {$q_0$};
        \node[circle, draw=black, right=of 1] (2) {$q_1$};
        \node[circle, draw=black, right=of 2] (3) {$q_2$};
        \node[circle, draw=gray, below=of 3] (4) {$q_3$};
        \node[circle, draw=black, left=of 4] (5) {$q_4$};
        \node[circle, draw=gray, left=of 5] (6) {$q_5$};
        \node[circle, draw=gray, below=of 6] (7) {$q_6$};
        \node[circle, draw=black, right=of 7] (8) {$q_7$};
        \node[circle, draw=black, right=of 8] (9) {$q_8$};
        \draw[->, thick, draw=black] (1) -- (2);
        \draw[thick, draw=gray] (1) -- (6);
        \draw[->, thick, draw=black] (2) -- (3);
        \draw[->, thick, draw=black] (2) -- (5);
        \draw[thick, draw=gray] (3) -- (4);
        \draw[thick, draw=gray] (4) -- (5);
        \draw[thick, draw=gray] (4) -- (9);
        \draw[thick, draw=gray] (5) -- (6);
        \draw[->, thick, draw=black] (5) -- (8);
        \draw[thick, draw=gray] (6) -- (7);
        \draw[thick, draw=gray] (7) -- (8);
        \draw[->, thick, draw=black] (8) -- (9);
    \end{tikzpicture}}
    & \qquad\qquad
	\left[
	\begin{array}{ccccccccc}
        1 & 1 & 0 & 0 & 0 & 0 & 0 & 0 & 0 \\
        0 & 1 & 0 & 0 & 0 & 0 & 0 & 0 & 0 \\
        0 & 0 & 1 & 0 & 0 & 0 & 0 & 0 & 0 \\
        0 & 0 & 0 & 1 & 0 & 0 & 0 & 0 & 0 \\
        0 & 0 & 0 & 0 & 1 & 0 & 0 & 1 & 0 \\
        0 & 0 & 0 & 0 & 0 & 1 & 0 & 0 & 0 \\
        0 & 0 & 0 & 0 & 0 & 0 & 1 & 0 & 0 \\
        0 & 0 & 0 & 0 & 0 & 0 & 0 & 1 & 0 \\
        0 & 0 & 0 & 0 & 0 & 0 & 0 & 0 & 1 \\
    \end{array}
    \right]
    \xrightarrow{\substack{R_1:=R_1\oplus R_2 \\ R_7:=R_7 \oplus R_8 \\ R_4:= R_4\oplus R_7 \\ R_1:=R_1\oplus R_4 \\ R_0:=R_0\oplus R_1}}
    \left[
	\begin{array}{ccccccccc}
        1 & 0 & 1 & 0 & 1 & 0 & 0 & 0 & 1 \\
        0 & 1 & 1 & 0 & 1 & 0 & 0 & 0 & 1 \\
        0 & 0 & 1 & 0 & 0 & 0 & 0 & 0 & 0 \\
        0 & 0 & 0 & 1 & 0 & 0 & 0 & 0 & 0 \\
        0 & 0 & 0 & 0 & 1 & 0 & 0 & 0 & 1 \\
        0 & 0 & 0 & 0 & 0 & 1 & 0 & 0 & 0 \\
        0 & 0 & 0 & 0 & 0 & 0 & 1 & 0 & 0 \\
        0 & 0 & 0 & 0 & 0 & 0 & 0 & 1 & 1 \\
        0 & 0 & 0 & 0 & 0 & 0 & 0 & 0 & 1 \\
    \end{array}
    \right]
\end{align*}
\endgroup
The first row of the final linear transformation above corresponds to the function $x_0 \mapsto x_0\oplus x_2 \oplus x_4 \oplus x_8$ as required.

Rather than uncompute the changes to the other rows, the \texttt{Gray-synth} algorithm takes the linear transformation into account when recursively partitioning the remaining parities. Once all parity computations have been completed, the algorithm computes the final linear transformation (see \cite{aam18}) using regular constrained Gaussian synthesis.

\subsection{Compilation}

Along with the default QASM output, \name includes a suite of source-to-source compilers or \emph{``transpilers"}, currently supporting output to Quil \cite{scz16}, ProjectQ \cite{sht18}, Q\# \cite{sgtaghkmpr18}, and Cirq \cite{cirq}. Effort has been made to translate QASM code to idiomatic code in the target language as much as possible -- in particular, translating \texttt{qelib1.inc} gates and gate declarations to standard library gates and idiomatic gate declarations in the target language whenever possible. \Cref{fig:trans} gives an example of the Q\# output for a QASM program.

\Name also includes an option to output just the resource counts of a program. By default the resource counter un-boxes resource counts for all declared gates except for those from the standard library, but the resource counter can be configured with a list of gates to leave boxed.

\subsection{Performance\label{sec:perf}}

To assess the performance of \name, we compare it against the well-known software toolkit and compiler Qiskit \cite{Qiskit}. \edit{}{We chose Qiskit to compare our work over the many other existing tools as it is arguably the largest compilation toolchain publicly available which supports the openQASM language.} In particular, we compare each tool's highest standard optimization setting and default hardware mapping settings for total gate counts and CNOT counts, respectively. In particular, we compare the Qiskit transpiler's level 3 optimization against \name's \texttt{-O2} command line option. The default mapping setting in \name applies the \texttt{steiner} mapping algorithm with the \texttt{best-fit} initial layout. For the optimization experiments, both tools unbox the program to the following subset of \texttt{qelib1.inc}:
\[
    \{ \texttt{u3}, \texttt{cx}, \texttt{h}, \texttt{rx}, \texttt{ry}, \texttt{rz} \}.
\]
We use a common benchmark suite \cite{amm14, nrscm17} to benchmark our compiler, consisting of largely reversible arithmetic and a few quantum algorithms. All experiments were run on 2.3GHz Intel Core i7 processor with 8GB of RAM running Arch Linux.

The results of optimization passes and mapping passes are reported in \Cref{tab:optimization,tab:mapping}, respectively. The best results in either case are identified in bold. For circuit optimization, \name beats Qiskit (over this gate set) for all but one of the benchmark circuits, achieving $31.7\%$ reduction in gate counts on average compared to Qiskit's $25.9\%$ average reduction. Similarly, in all but one benchmark circuit with the highest number of qubits, \name was also significantly faster. It remains a focus of future work to improve the scalability of \name's optimization algorithms as the number of qubits increases.

For hardware mapping benchmarks, IBM's $20$ qubit Tokyo chip was selected as the target architecture, and as such only the benchmark circuits which fit onto the chip are reported in \Cref{tab:mapping}. In contrast to optimization, the experimental results for hardware mapping were mixed. While the \texttt{Steiner} mapping algorithm with \texttt{best-fit} initial layout was consistently orders of magnitude faster than Qiskit's default mapping algorithm, Qiskit outperformed \name in terms of CNOT counts in the majority of cases. As hardware mapping is very sensitive to initial qubit placement \cite{p19}, similar to \cite{km19} a simple \emph{hill-climb} algorithm was implemented to optimize the initial layout and combined with the \texttt{Steiner} mapping algorithm (last two columns of \Cref{tab:mapping}). With this qubit layout optimization, \name's default mapping algorithm outperforms Qiskit in the majority of cases, with an average CNOT-count increase of $103.4\%$ compared to Qiskit's $125.7\%$, while still running faster than Qiskit in almost all cases. Moreover, many of the cases where Qiskit outperformed \name appear to be pathological cases for the underlying \texttt{Gray-synth} algorithm \cite{aam18}. It remains to be seen whether more sophisticated layout optimization algorithms which avoid local minima -- for instance, simulated annealing -- can improve the results further.

\begin{table}[]
    \centering
    \small
    \begin{tabular}{lrrrrrrrr}
        \toprule
        Benchmark & $n$ & \multicolumn{2}{c}{Original} & \multicolumn{2}{c}{Qiskit} & \multicolumn{3}{c}{\Name} \\ 
        \cmidrule(lr){3-4} \cmidrule(lr){5-6} \cmidrule(lr){7-9}
        & & \# gates & depth & \# gates & time (s) & \# gates & depth & time (s) \\ \midrule
Grover$\_5$&9&1023&320&769&9.64&\bf 673&203&0.203 \\
Mod $5\_4$&5&79&59&60&2.605&\bf 53&37&0.029 \\
VBE-Adder$\_3$&10&190&63&134&3.421&\bf 105&38&0.031 \\
CSLA-MUX$\_3$&15&210&37&160&3.638&\bf 158&28&0.039 \\
CSUM-MUX$\_9$&30&532&38&420&6.132&\bf 336&26&0.091 \\ 
QCLA-Com$\_7$&24&659&81&416&6.287&\bf 341&51&0.095 \\
QCLA-Mod$\_7$&26&1120&137&838&10.888&\bf 740&97&0.246 \\
QCLA-Adder$\_10$&36&657&61&494&7.069&\bf 443&40&0.128 \\
Adder$\_{8}$&24&1128&157&869&11.485&\bf 751&113&0.254 \\
RC-Adder$\_{6}$&14&244&45&185&3.977&\bf 173&34&0.055 \\
Mod-Red$\_{21}$&11&346&87&261&4.618&\bf 241&60&0.055 \\
Mod-Mult$\_{55}$&9&147&37&117&3.146&\bf 109&29&0.036 \\
Mod-Adder$\_{1024}$&28&5425&1188&3871&52.066&\bf 3335&793&1.039 \\
GF($2^4$)-Mult&12&289&54&213&4.256&\bf 203&37&0.079 \\
GF($2^5$)-Mult&15&447&66&327&5.322&\bf 317&44&0.097 \\
GF($2^6$)-Mult&18&639&78&\bf 357&7.039&444&52&0.163 \\
GF($2^7$)-Mult&21&865&90&627&8.906&\bf 606&59&0.29 \\
GF($2^8$)-Mult&24&1139&106&819&11.018&\bf 791&71&0.513 \\
GF($2^9$)-Mult &27&1419&114&1023&13.551&\bf 987&74&0.782 \\
GF($2^{10}$)-Mult&30&1747&126&1257&16.386&\bf 1202&82&1.15 \\
GF($2^{16}$)-Mult&48&4459&202&3179&42.435&\bf 3059&131&10.324 \\
GF($2^{32}$)-Mult&96&17658&396&12536&4m&\bf 12042&253&5m \\
Ham$\_{15}$ (low)&17&535&165&425&6.689&\bf 391&108&0.111 \\
Ham$\_{15}$ (med)&17&1599&528&1154&15.027&\bf 994&344&0.241 \\
Ham$\_{15}$ (high)&20&6712&2244&4768&1m&\bf 3982&1448&1.395 \\
HWB$\_6$&7&319&93&248&16.151&\bf 232&68&0.056 \\
HWB$\_8$&12&18220&3102&14059&5m&\bf 12827&2257&2.939 \\
QFT$\_4$&5&187&117&161&3.927&\bf 178&117&0.040 \\
$\Lambda_3(X)$&5&57&34&44&2.531&\bf 40&26&0.027 \\
$\Lambda_3(X)$ (Barenco)&5&76&40&56&2.955&\bf 50&32&0.024 \\
$\Lambda_4(X)$&7&95&40&73&2.92&\bf 72&28&0.028 \\
$\Lambda_4(X)$ (Barenco)&7&146&64&109&3.178&\bf 98&44&0.034 \\
$\Lambda_5(X)$&9&133&40&101&3.148&\bf 90&28&0.035 \\
$\Lambda_5(X)$ (Barenco)&9&218&76&162&3.792&\bf 146&54&0.035 \\
$\Lambda_{10}(X)$&19&323&40&247&4.741&\bf 215&28&0.05 \\
$\Lambda_{10}(X)$ (Barenco)&19&578&76&427&6.545&\bf 386&54&0.077 \\
\midrule
Average reduction (\%) & & & & 25.9 & & \bf 31.7 & &  \\ \bottomrule
    \end{tabular}
    \caption{Benchmark optimization results}
    \label{tab:optimization}
\end{table}

\begin{table}[]
    \centering
    \small
    \begin{tabular}{lrrrrrrrr}
        \toprule
        Benchmark & $n$ & Original & \multicolumn{2}{c}{Qiskit} & \multicolumn{2}{c}{\Name} & \multicolumn{2}{c}{\Name} \\
        & & & & & & & \multicolumn{2}{c}{(w/ layout opt.)} \\ 
        \cmidrule(lr){3-3} \cmidrule(lr){4-5} \cmidrule(lr){6-7} \cmidrule(lr){8-9}
        & & \# CNOTs & \# CNOTs & time (s) & \# CNOTs & time (s) & \# CNOTs & time (s) \\ \midrule
Grover$\_5$&9&288&703&31.226&592&0.204&\bf 422&5.763 \\
Mod $5\_4$&5&28&58&4.040&43&0.037&\bf 40&0.084 \\
VBE-Adder$\_3$&10&70&134&7.770&78&0.039&\bf 68&0.968 \\
CSLA-MUX$\_3$&15&80&208&8.419&240&0.055&\bf 177&5.202 \\
RC-Adder$\_{6}$&14&93&239&9.307&\bf 150&0.071&151&3.514 \\
Mod-Red$\_{21}$&11&105&243&10.766&280&0.081&\bf 187&2.777 \\
Mod-Mult$\_{55}$&9&48&123&6.013&142&0.043&\bf 109&1.01 \\
GF($2^4$)-Mult&12&99&\bf 284&10.483&422&0.077&337&2.495 \\
GF($2^5$)-Mult&15&154&\bf 472&17.922&589&0.126&478&6.599 \\
GF($2^6$)-Mult&18&221&\bf 749&26.492&851&0.196&765&15.847 \\
Ham$\_{15}$ (low)&17&236&\bf 794&26.806&887&0.146&887&3.605 \\
Ham$\_{15}$ (med)&17&534&1630&1m&\bf 1406&0.330&\bf 1406&9.934 \\
Ham$\_{15}$ (high)&20&2149&5852&4m&\bf 5512&1.466&\bf 5512&55.176 \\
HWB$\_6$&7&116&\bf 237&11.427&259&0.074&259&0.430 \\
HWB$\_8$&12&7129&19650&9m&22687&4.303&\bf 17428&5m \\
QFT$\_4$&5&46&\bf 72&6.271&100&0.058&90&0.284 \\
$\Lambda_3(X)$&5&18&\bf 18&3.157&32&0.027&30&0.091 \\
$\Lambda_3(X)$ (Barenco)&5&24&\bf 24&3.446&35&0.037&35&0.076 \\
$\Lambda_4(X)$&7&30&\bf 43&4.218&61&0.035&47&0.359 \\
$\Lambda_4(X)$ (Barenco)&7&48&\bf 48&3.351&73&0.036&58&0.417 \\
$\Lambda_5(X)$&9&42&86&5.448&101&0.036&\bf 70&1.041 \\
$\Lambda_5(X)$ (Barenco)&9&72&126&6.469&150&0.046&\bf 102&1.654 \\
$\Lambda_{10}(X)$&19&102&224&12.916&238&0.078&\bf 134&24.583 \\
$\Lambda_{10}(X)$ (Barenco)&19&192&470&13.832&404&0.117&\bf 231&26.959 \\
    \midrule
    Average increase (\%) & & & 125.7 & & 146.4 & & \bf 103.4 &  \\ \bottomrule
    \end{tabular}
    \caption{Hardware mapping results}
    \label{tab:mapping}
\end{table}

\section{Conclusions and future directions\label{sct:concl}}
In this article we described \name along with its main use cases, and provided a set of benchmarks. \Name is a modular quantum compiling toolkit, which is easy to extend, its design being inspired by Clang~\cite{clang}.

The dynamic field of quantum software is evolving rapidly, being driven by a variety of factors, ranging from progress in quantum hardware to improved compilation techniques. While we cannot foresee what the future will reserve, we still have a set of QASM-based features we will most likely add to \name, such as: i) more QASM syntax extensions that will not break backwards compatibility, ii) ability to perform iterations and loops, iii) having registers as arguments to gates instead of qubits. Such extensions will allow the user to design quantum software libraries in a relatively straight-forward manner while focusing on efficiently achieving the desired functionality.

\bibliographystyle{unsrt}


\appendix
\section{Transpiler example}

\begin{figure}
\centering
\begin{subfigure}[b]{1\textwidth}
\centering
\begin{lstlisting}[basicstyle=\ttfamily\tiny]
OPENQASM 2.0;
include "qelib1.inc";
 
gate bell x,y {
  h x;
  cx x,y;
}

qreg q[1];
qreg anc[2];
creg c0[1];
creg c1[1];

bell anc[0],anc[1];
cx q,anc[0];
h q;
measure q -> c0[0];
measure anc[0] -> c1[0];
if(c0==1) z anc[1];
if(c1==1) x anc[1];
\end{lstlisting}
\caption{QASM source}
\end{subfigure}
\begin{subfigure}[b]{1\textwidth}
\centering
\begin{lstlisting}[basicstyle=\ttfamily\tiny]
namespace Quantum.staq {
    open Microsoft.Quantum.Intrinsic;
    open Microsoft.Quantum.Convert;
    open Microsoft.Quantum.Canon;
    open Microsoft.Quantum.Math;

    operation U(theta : Double, phi : Double, lambda : Double, q : Qubit) : Unit {
        Rz(lambda, q);
        Ry(theta, q);
        Rz(phi, q);
    }
    operation u3(theta : Double, phi : Double, lambda : Double, q : Qubit) : Unit {
        U(theta, phi, lambda, q);
    }
    operation u2(phi : Double, lambda : Double, q : Qubit) : Unit {
        U(PI()/2.0, phi, lambda, q);
    }
    operation u0(gamma : Double, q : Qubit) : Unit {
        U(0.0, 0.0, 0.0, q);
    }
    operation cy(a : Qubit, b : Qubit) : Unit {
        (Adjoint S)(b);
        CNOT(a, b);
        S(b);
    }
    operation swap(a : Qubit, b : Qubit) : Unit {
        CNOT(a, b);
        CNOT(b, a);
        CNOT(a, b);
    }
    operation cu3(theta : Double, phi : Double, lambda : Double, c : Qubit, t : Qubit) : Unit {
        Rz((lambda-phi)/2.0, t);
        CNOT(c, t);
        u3(-(theta/2.0), 0.0, -((phi+lambda)/2.0), t);
        CNOT(c, t);
        u3(theta/2.0, phi, 0.0, t);
    }

    operation bell(x : Qubit, y : Qubit) : Unit {
        H(x);
        CNOT(x, y);
    }

    operation Circuit() : Unit {
        using (q = Qubit[1]) {
            using (anc = Qubit[2]) {
                mutable c0 = new Result[1];
                mutable c1 = new Result[1];
                bell(anc[0], anc[1]);
                CNOT(q[0], anc[0]);
                H(q[0]);
                set c0 w/= 0 <- M(q[0]);
                set c1 w/= 0 <- M(anc[0]);
                if (ResultArrayAsInt(c0) == 1) {
                    Z(anc[1]);
                }
                if (ResultArrayAsInt(c1) == 1) {
                    X(anc[1]);
                }

                ResetAll(anc);
            }
            ResetAll(q);
        }
    }
}
\end{lstlisting}
\caption{Translated Q\# code}
\end{subfigure}
\caption{Translation between QASM and Q\#}
\label{fig:trans}
\end{figure}

\end{document}